\documentclass[fleqn,usenatbib]{mnras}
\usepackage{newtxtext,newtxmath}
\usepackage[T1]{fontenc}
\usepackage{ae,aecompl}
\usepackage{xspace}

\usepackage{graphicx}	% Including figure files
\usepackage{amsmath}	% Advanced maths commands
\usepackage{amssymb}	% Extra maths symbols
\usepackage{pdflscape}
\usepackage{tabu}
\usepackage{subfig}
\usepackage[para,online,flushleft]{threeparttable}
\usepackage{ulem}

\makeatletter
\newcommand*{\rom}[1]{\expandafter\@slowromancap\romannumeral #1@}
\makeatother

\newcommand{\rrata}{J0621$-$55\xspace}
\newcommand{\rratb}{J1646$-$1910\xspace}
\newcommand{\nuller}{J1337$-$4441\xspace}
\newcommand{\troll}{J1306$-$4035\xspace}
\newcommand{\twelve}{J2251$-$3711\xspace}
\defcitealias{kbj+18}{Paper~I}
\newcommand{\paperi}{\citetalias{kbj+18}\xspace}
\newcommand{\snr}{\ensuremath{S/N}\xspace}

\title[SUPERB V]{The SUrvey for Pulsars and Extragalactic Radio Bursts V: \\ 
Recent Discoveries and Full Timing Solutions}

\author[R.~Spiewak et al.]{
R.~Spiewak,$^{1,2}$\thanks{\href{mailto:rspiewak@swin.edu.au}{rspiewak@swin.edu.au (RS)}}
C.~Flynn,$^1$
S.~Johnston,$^3$
E.~F.~Keane,$^{4}$
M.~Bailes,$^{1,2}$
E.~D.~Barr,$^5$  \newauthor
S.~Bhandari,$^3$
M.~Burgay,$^6$ 
F.~Jankowski,$^7$
M.~Kramer,$^{5,7}$
V.~Morello,$^7$
A.~Possenti,$^{6,8}$  \newauthor
V.~Venkatraman Krishnan$^5$
\\
$^{1}$ Centre for Astrophysics and Supercomputing, Swinburne University of Technology, PO Box 218, Hawthorn, VIC 3122, Australia\\
$^{2}$ ARC Centre of Excellence for Gravitational Wave Discovery (OzGrav), Mail H29, Swinburne University of Technology, PO Box 218, Hawthorn, \\ VIC 3122, Australia \\ 
$^3$ CSIRO Astronomy and Space Science, Australia Telescope National Facility, PO Box 76, Epping, NSW 1710, Australia \\
$^4$ SKA Organisation, Jodrell Bank, Macclesfield SK11 9FT, UK \\
$^5$ Max-Planck-Institut f\"ur Radioastronomie, Auf dem H\"ugel 69, D-53121 Bonn, Germany \\
$^6$ INAF-Osservatorio Astronomico di Cagliari, via della Scienza 5, 09047 Selargius (CA), Italy \\
$^7$ Jodrell Bank Centre for Astrophysics, Department of Physics and Astronomy, The University of Manchester, Manchester M13 9PL, UK \\
$^8$ Universit\`a di Cagliari, Dip di Fisica, S.P. Monserrato-Sestu Km 0,700 I-09042 Monserrato, Italy 
}

\date{Accepted XXX. Received YYY; in original form ZZZ}

\pubyear{2020}

\begin{document}
\label{firstpage}
\pagerange{\pageref{firstpage}--\pageref{lastpage}}
\maketitle

% Abstract of the paper
\begin{abstract}
The SUrvey for Pulsars and Extragalactic Radio Bursts ran from 2014 April to 2019 August, covering a large fraction of the southern hemisphere at mid- to high-galactic latitudes, and consisting of 9-minute pointings taken with the 20-cm multibeam receiver on the Parkes Radio Telescope. 
Data up to 2017 September 21 have been searched using standard Fourier techniques, single-pulse searches, and Fast Folding Algorithm searches. 
We present 19 new discoveries, bringing the total to 27 discoveries in the programme, and we report the results of follow-up timing observations at Parkes for 26 of these pulsars, including the millisecond pulsar PSR~J1421$-$4409; the faint, highly-modulated, slow pulsar PSR~J1646$-$1910; and the nulling pulsar PSR~J1337$-$4441. 
We present new timing solutions for 23 pulsars, and we report flux densities, modulation indices, and polarization properties. 

\end{abstract}

\begin{keywords}
pulsars: general -- surveys -- methods: data analysis -- methods: observational
\end{keywords}

%%%%%%%%%%%%%%%%%%%%%%%%%%%%%%%%%%%%%%%%%%%%%%%%%%

%%%%%%%%%%%%%%%%% BODY OF PAPER %%%%%%%%%%%%%%%%%%

\section{Introduction}
\label{sec:superb_intro}

The Parkes Radio Telescope has been a prolific instrument in the search for pulsars over the decades. 
The Parkes southern Sky Survey was conducted at Parkes using a 70-cm receiver and discovered 101 pulsars \citep[see, e.g.,][]{bhl+94,lnl+95} including the bright, nearby millisecond pulsar (MSP), PSR~J0437$-$4715 \citep{jlh+93}. 
With the advent of the 13-beam receiver at Parkes \citep{sswb+96} and large bandwidth analogue filterbanks, the Parkes Multibeam Pulsar Survey \citep{mlc+01} was able, between 1998 and 2003, to search a wide unexplored parameter space and discovered over 800 pulsars\footnote{From the ATNF pulsar catalogue (\url{https://www.atnf.csiro.au/research/pulsar/psrcat/}), including from re-processing; e.g., \citealt{mmemll18}.} within 5 degrees of the Galactic plane \citep{mhl+02,kbm+03,hfs+04,fsk+04,lfl+06}. 
Later, the High Time Resolution Universe (HTRU-S) survey, carried out between 2008 and 2014, employed higher time and frequency resolution in an all-sky survey with the Parkes multibeam to discover over 200 further pulsars and 10 Fast Radio Bursts \citep[FRBs;][]{kjvs+10,bbb+11b,bsbj+11,tsb+13,cpk+16,pos+19}.

HTRU-S was directly followed by another major southern hemisphere survey, SUPERB ``The SUrvey for Pulsars and Extragalactic Radio Bursts'' \citep[][hereafter Paper~I]{kbj+18}. 
SUPERB also used the Parkes multibeam receiver, and observations were conducted between April 2014 and August 2019. 
The primary goals of SUPERB included the real-time discovery of FRBs and pulsars and improving upon previous surveys in the region. 
SUPERB observed each pointing for 9.3\,minutes, compared with 4.5\,minutes for the high-latitude (``hilat'') portion of HTRU-S \citep{kjvs+10}, and the pointings were tessellated such that the most sensitive central beam of the instrument was placed at positions of the less sensitive ``outer-ring'' pointings of HTRU-S. 
The pointings were additionally offset by half a half-power beamwidth so that some points previously at the half-power point in HTRU-S hilat observations were on-axis for SUPERB and vice-versa. 
Finally, SUPERB increased the chances of detecting intermittent pulsars and RRATs by observing pointings multiple times (the actual number of observations per position varied widely, averaging $\sim2.3$ observations per position excluding singly-observed pointings).

Throughout the last two decades, in the era of the multibeam surveys at Parkes, the radio frequency interference (RFI) environment has worsened with time. 
Progress is made in this (and other) regards by repeated re-processing of these archival data as new RFI mitigation and search algorithms are developed. 
This has been done successfully in the past \citep{ekl09,kel+09,kek+13,mbb+14,mbc+19} and we expect the same will be true for SUPERB.
Decreasing the discovery lag for pulsars is an important step for enabling large-volume surveys such as will be performed with the Square Kilometer Array (SKA; \citealt{kbks+14}). 
In addition, finding pulsars in real-time is crucial for studies of intermittent pulsars and RRATs, which may be visible for only a short time and benefit from rapid follow-up for confirmation and immediate study.

\paperi described the survey parameters in detail and presented the first pulsars discovered in the survey until January 2016\footnote{\url{https://sites.google.com/site/publicsuperb/discoveries}}. 
The 6 FRBs discovered to date in SUPERB data were described by \citet{kjb+16}, \citet{bkb+18}, and \citep{bck+18}, and their polarization properties in \citet{cks+18}. 
\citet{mke+20} presented the discovery of a particularly noteworthy pulsar with a spin period of $\sim$12 seconds, PSR~\twelve, as it raises interesting questions concerning the diversity of the pulsar population and pulsar evolution and
emission models.

In this paper, we report the discovery of an additional 19 pulsars, bringing the total number of SUPERB discoveries to 27, and present new timing solutions for 23 of these pulsars.
Our sample includes a nulling pulsar, PSR~\nuller; an MSP in a wide, eccentric orbit, PSR~J1421$-$4409; and a faint, highly-modulated 4.8-second pulsar, PSR~J1646$-$1910, which exhibits RRAT-like emission. 
A second MSP, PSR~J1306$-$4035, remains unsolved due to insufficient detections, possibly due to eclipsing in this redback system. 
We further present integrated profiles together with polarization properties and compare the emission properties of our sample with the known Galactic population. 

In Section~\ref{ssec:superb_obs}, we describe the methodology used to collect the pulsar data and present our discoveries. 
In Section~\ref{ssec:superb_data}, we report the results of the follow-up timing observations including the timing parameters, the flux density measurements and the effects of scintillation, and the polarization properties of the pulsars. 
We discuss in more detail the most noteworthy pulsars in our sample in Section~\ref{sec:superb_rrats}, and conclude in Section~\ref{sec:superb_disc}.

%%%%%%%%%%% Section 2
\section{Observations and Data Analysis}

\subsection{Survey and Pulsar Follow-up Observations}
\label{ssec:superb_obs}

\paperi describes the overall set-up of the survey; here, we summarise those aspects most relevant to the results of our analysis. 
In this publication, we consider survey data collected up to 2017 September 21. 
22,007 observations (285,983 independent pointings) were performed up to that date, of which 21,182 observations (275,296 pointings) have been thoroughly searched and inspected for pulsar candidates, including follow-up observations of various sources. 
After accounting for repeated pointings, a total of 224,899 sky positions were observed up to 2017 September 21 ($\approx84$\% of planned pointings according to \paperi). 
After the cut-off date, processing is not complete due in part to the migration from the gSTAR supercomputer to the new OzSTAR\footnote{\url{http://supercomputing.swin.edu.au/}} cluster at Swinburne University of Technology.

\begin{figure*}
    \centering
    \includegraphics[width=\textwidth]{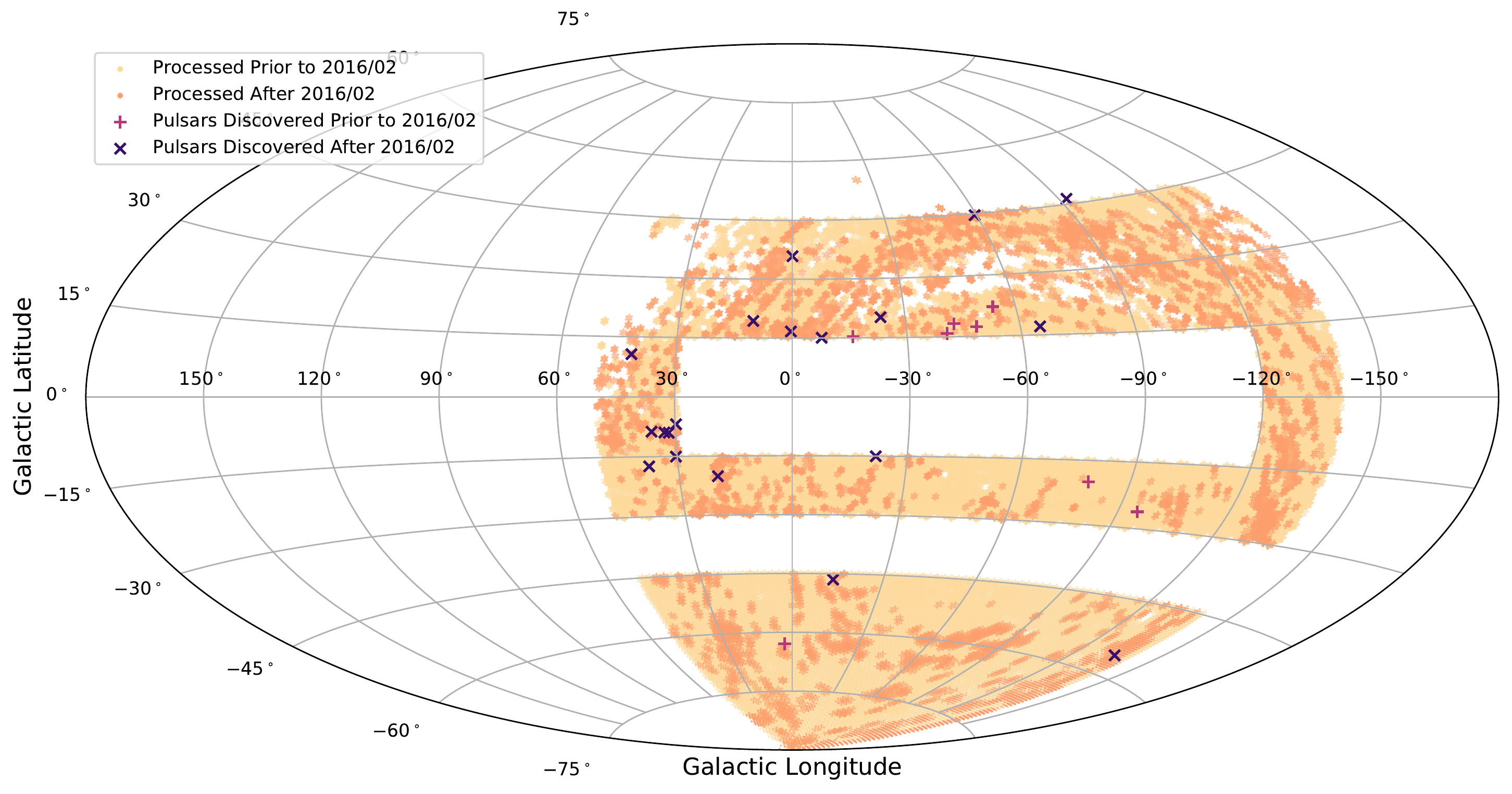} % high-res % too high-res
    \caption[SUPERB sky coverage]{The sky covered by the SUPERB survey until 2017 September. The observed beams (including any follow-up observations) that were processed using the search pipelines described in the text are plotted in light orange (processed prior to 2016 January) and orange, and the pulsars discovered prior to (after) 2016 January are plotted as purple `+' (dark blue `x') marks.}
    \label{fig:superb_sky}
\end{figure*}

\begin{figure}
    \centering
    \includegraphics[width=0.475\textwidth]{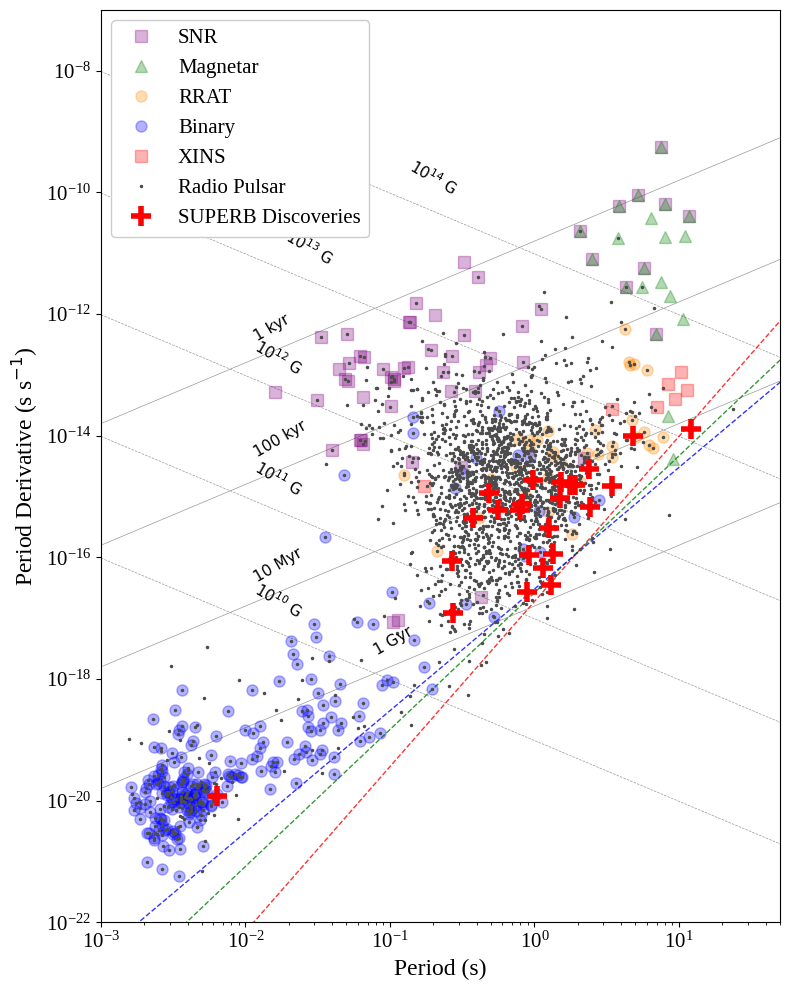}
    \caption[$P-\dot P$ diagram highlighting SUPERB pulsars]{$P-\dot{P}$ diagram, as in \citet{mke+20}, based on v1.63 of the ATNF pulsar catalogue \citep{mhth05}. The pulsars discovered in SUPERB, including PSR~\twelve, have been highlighted in red. Lines of constant characteristic age and surface magnetic field strength are displayed in grey. The dashed red line represents the lower limit of the so-called pulsar death valley \citep[Eq. 8 of][]{cr93a}. Death lines from \citet{zhm00} are also shown, based on their curvature radiation from vacuum gap model (green dashes) and space-charged-limited flow model (blue dashes).}
    \label{fig:superb_ppdot}
\end{figure}

SUPERB observations (Parkes observing programmes P858 and P892, PI Keane) cover the survey region shown in Fig.~\ref{fig:superb_sky}, covering most of the sky with $b < -45^{\circ}$ or $-30^{\circ} < b < 45^{\circ}$, and $-140^{\circ} < l < 50^{\circ}$, and excluding the region covered by the intermediate latitude part of HTRU-S. 
Of the 27 pulsars discovered by SUPERB through 2017 September, 24 have full timing solutions with significant period derivatives and are highlighted in Fig.~\ref{fig:superb_ppdot}.
The timing programme described herein focused on the 26 pulsars excluding PSR~\twelve. 

We note that four of the pulsars included in this work were previously published in \paperi, with different names based on their discovery parameters. We list these here for reference: \begin{itemize}
    \item PSR~J0750$-$6846 -- previously J0749$-$68
    \item PSR~J1337$-$4441 -- previously J1337$-$44
    \item PSR~J1406$-$4233 -- previously J1405$-$42
    \item PSR~J1604$-$3142 -- previously J1604$-$31
\end{itemize}

Standard SUPERB observations use only the Berkeley-Parkes-Swinburne-Recorder (BPSR) with the HI-Pulsar Signal Recorder (HIPSR; \citealt{pssb+16}) to record data at 64\,$\upmu$s time resolution and 0.78\,MHz frequency resolution. 
To achieve higher phase resolution, follow-up timing observations were made with the CASPER Parkes Swinburne Recorder\footnote{\url{http://astronomy.swin.edu.au/pulsar/?topic=caspsr}} (CASPSR) and the Digital Filterbank Mark \rom{4} (DFB4), using the discovery DM and period to fold the data, with frequency resolutions of 0.78 and 0.5\,MHz for CASPSR and DFB4, respectively. 
These back-ends are described in Table~\ref{tab:superb_obs} \citep[see also][]{scvs11}. 
We additionally recorded data from all 13 beams using the BPSR backend while observing the pulsars, enabling further searches for FRBs and pulsars simultaneously. 
Typical survey pointings are $\sim$560\,s, while observations of known FRB positions, to search for possible repeats, vary in length up to a few hours. 
PSRs~J1921$-$0510 and J1923$-$0408 were discovered in outer beams of such follow-up observations. 
PSR~J1923$-$0408 was additionally detected (at very low significance) in the original 560-second observation at that position, and PSR~J1921$-$0510 was detected (not in a blind search) in an earlier 560-second observation.

\begin{table}
 \centering 
 \caption[SUPERB Parkes observation information]{Observations -- relevant information for the receivers used in this project. \label{tab:superb_obs}}
 \begin{scriptsize}
 \begin{tabular}{l l c c l}
  \hline
  Receiver & Backend & Centre Freq. & Bandwidth & Dates \\ 
           &         & (MHz)        & (MHz)     & (MJD) \\ \hline
  MB       & CASPSR  & 1382         & 340       & 58044-58405 \\
           & BPSR    & 1382         & 340       & 57116-58405 \\
           & DFB4    & 1369         & 256       & 58044-58405 \\ 
  UWL      & Medusa  & 2350         & 3300      & 58468 \\ 
           & CASPSR  & 724    & 64  & 58468 \\
           & DFB4    & 1369         & 256       & 58468 \\ \hline
 \end{tabular}
 \end{scriptsize}
\end{table}

Using the Green II supercomputing cluster located at Swinburne University, the observations were automatically processed through a set of pipelines for detecting FRBs and pulsars. 
As described in \paperi, the main pipelines use a periodicity search and a single-pulse detector. 
The output of these pipelines from the data through to 2016 February was published in \paperi. 
Additionally, an experimental, large-scale Fast Folding Algorithm (FFA; \citealt{staelin69}) search was tested on the SUPERB data, revealing 11 additional pulsars in the data through 2016 February. 
This pipeline was not fully developed at the time and was run on only a subset of the data; completion of the pipeline and search is anticipated in the near future \citep[][submitted]{mbskl20}. 
All promising candidates identified by the various pipelines were allocated time (as part of the survey programme) for confirmations as soon after discovery as practicable. 
Those that were detected in multiple observations were thereafter considered confirmed discoveries, after checking the public lists of discoveries by other ongoing surveys. 
Follow-up observations for this paper targeted all SUPERB discoveries, excluding PSR~J1421$-$4409 which was observed substantially in the MSP and Binary Pulsar observing programme at Parkes, P789 (PI Spiewak).

Timing observations, as part of the P892 observing programme (PI Keane), began in 2017 October but were ended early in 2018 October when the Multibeam receiver was removed from the Parkes focus cabin to enable observations of Voyager 2 as it passed through the heliopause \citep{rbggb19}. 
In 2018 December, we requested and were allocated 4 hours of Parkes Director's Time to observe 8 pulsars with the Ultra-Wideband Low (UWL; \citealt{hmd+19}) receiver using the Medusa backend, with the intention of further constraining pulsar dispersion measures (DMs) and spectral indices\footnote{UWL observations, in fold-mode, were made at 1.0\,MHz frequency resolution on 2018 December 16.}. 
For the five pulsars successfully detected with the UWL, the constraints on DM were consistent with those from timing using the 20-cm data (using sub-banded times of arrival (ToAs)), and spectral indices were not significantly constrained, primarily due to low \snr biasing the detections to the $\sim1$\,--\,2\,GHz range.

Although most of the pulsars discovered in this survey were originally detected in 560-second pointings, the integration times for follow-up timing have varied from 5 to 60\,minutes, depending on the brightness and emission/scintillation timescale of each source. 
In total, we performed approximately 190\,hours of follow-up timing in the P892 programme for the 26 sources in this paper (including two observations of PSR~J1421$-$4409). 

In order to improve the positions of some pulsars (and thereby reduce the integration time required to detect the pulsar signal), we performed ``gridding'' observations \citep[see, e.g.,][]{mhl+02} of all pulsars requiring less than $\sim30$\,minutes of integration time. 
Our procedure was simply to offset the central beam $10\arcmin$ (beyond the $7\arcmin$ half-power point for the beam, but less than the $29\arcmin$ beam separation) in Right Ascension and Declination, for a total of 5 pointings including the initial position. 
The CASPSR data were then reduced in the same manner as the regular timing observations (summarised below), and the resulting \snr values were fit by a 2-dimensional Gaussian with a full-width-half-maximum (FWHM) of $14\arcmin$ matching that of the central beam of the Parkes Multibeam receiver. 
Due to scintillation or low \snr detections, the positions derived for most pulsars had large uncertainties and those for only 3 pulsars were used in further observations. 
These positions were, in fact, closer to the true positions (found through timing) than the discovery positions, with improvements of 1.2\,--\,4.1\,arcmin ($\approx$\,10\,--\,30\,percent of the beam FWHM) in position accuracy.

In addition to these observations in the P892 observing programme, the MSPs, PSRs~\troll and J1421$-$4409, were added to the P789 observing programme. 
The observations of PSR~J1421$-$4409 from this programme have been used to update the timing solution from \paperi (see Section~\ref{ssec:superb_1421}).

\subsection{Data Analysis}
\label{ssec:superb_data}
% sec 2.2 

The processing of the pulsar timing data, using primarily the coherently de-dispersed CASPSR data, was done via a script to ensure consistency and efficiency. 
The script used system calls to various \textsc{psrchive} routines\footnote{\url{http://psrchive.sourceforge.net/}} \citep{vsdo12},
including \textsc{psrsh} commands for RFI excision, \textsc{pac} for polarization and flux density calibration, and \textsc{psrflux} for calculation of flux density values. 
We additionally observed the standard noise diode for polarization calibration and used observations of Hydra A (primarily from other Parkes projects, such as P456, with permission) for flux density calibration. 
From the processed data, analytic templates were created from the brightest observation, or the sum of the 5 brightest observations for fainter sources. 
We formed pulse times of arrival (ToAs) using \textsc{pat} with the Fourier domain with Markov Chain Monte Carlo algorithm \citep{vanstraten06}, 
after fully integrating the frequency and polarization channels, with sub-integration lengths varying between 1 and 15 minutes (optimised for \snr per sub-integration). 
We additionally used sub-banded ToAs (typically with 4 frequency channels per sub-integration) to verify DM values. 
We used the \textsc{tempo2} software package\footnote{\url{https://bitbucket.org/psrsoft/tempo2/src/master/}} \citep{hem06} for all timing and used Barycentric Coordinate Time (TCB) units and the DE421 planetary ephemeris model \citep{fwb09}.

For most of the pulsars, observations with BPSR prior to 2017 October were available and suitable for inclusion in the timing, serving to increase the time span. 
These data were first folded using \textsc{dspsr} \citep{vb11} before being processed in the same manner as described above. 
Templates generated from CASPSR data were used to measure ToAs from the BPSR data. 
The fitting of timing models had to account for phase jumps between different back-ends, which were not fixed to known values but fit using routine simultaneous observations. 
We also processed DFB4 data where necessary to replace missing CASPSR observations (as these were recorded simultaneously when possible), and polarization-calibrated DFB4 data were used to produce the pulsar profiles, which are shown in Figures~\ref{fig:superb_polsA}--\ref{fig:superb_polsE}. 
Where possible, CASPSR data were preferred over DFB4 data for the timing analyses due to the coherent de-dispersion (which improves the time resolution for MSPs like PSR~J1421$-$4409) and shorter sub-integrations in the raw data (allowing more precise excision of RFI).

The timing solutions and additional parameters, including polarization properties and flux density measurements, are described in Tables~\ref{tab:superb_time} and \ref{tab:superb_derv}. 
We measured the flux density of each sub-integration (frequency-integrated to the central frequency of $\approx$1382\,MHz) and analysed the distributions to determine the median flux density, $S_{1400}$, in addition to the modulation indices, $m=\sigma_{S1400}/\Bar{S}_{1400}$ (using the mean flux density, not the median value). 
All flux density values have been corrected for position offsets, which vary from 2 to 8\,arcmin, using a Gaussian beam correction \citep[see, e.g., ][]{rkp+11}. 
We note that the sub-integrations varied in length from 1 to 15 minutes, affecting our sampling of the scintillation structure, and that all pulsars in our sample should be in the ``strong'' regime of scintillation. 
All flux density distributions were limited to exclude sub-integrations where $S_{1400} < 2\sigma_{S1400}$ except PSR~\nuller, for which we excluded $S_{1400} < 3\sigma_{S1400}$ to exclude the nulling behaviour.

\begin{table*}
    %\centering
    \caption[Timing parameters for SUPERB pulsars]{Timing parameters for all (solved) pulsars, excluding PSR~J1421$-$4409 (see Table~\ref{tab:superb_j1421}). 
    All uncertainties are quoted at the 1-$\sigma$ level from the \textsc{tempo2} fits to the data. 
    The names of the newly discovered pulsars are marked with asterisks. 
    }
    \label{tab:superb_time}
    \begin{scriptsize}
    \begin{tabular}{l*{9}{c}} \hline 
         Pulsar       & Right            & Declination       & Period             & Period      & Reference & Data & r.m.s.     & Number  \\ 
                      & Ascension        &                   &                    & Derivative  & Epoch     & Span & (Weighted) & of ToAs  \\
                      & (J2000; h:m:s) & (J2000; d:m:s) & ($P$; s) & ($\dot P$; $10^{-15}$\,s\,s$^{-1}$) & (MJD)  & (yr) & (ms)  &       \\ \hline
         J0750$-$6846 & 07:50:35.85(2)   & $-$68:46:33.81(10)  & 0.91521580361(7)   & 0.1096(7)   & 56839  & 3.41 & 1.33  & 85    \\ %0
         J1115$-$0956* & 11:15:27.361(5)  & $-$09:56:53.1(2)    & 1.31089402030(2)   & 0.0347(3)   & 57317  & 2.94 & 0.527 & 54    \\ %1
         J1207$-$4508* & 12:07:19.26(2)   & $-$45:08:09.1(4)    & 1.769827846(2)     & 1.56(2)     & 56936  & 1.03 & 1.59  & 186   \\ %2
         J1244$-$1812* & 12:44:25.78(5)   & $-$18:12:16.2(9)    & 3.4253237068(2)    & 1.483(12)   & 57828  & 1.61 & 1.43  & 46    \\ %3
         J1337$-$4441 & 13:37:27.749(10) & $-$44:41:42.3(2)    & 1.25751149260(3)   & 0.3031(5)   & 56933  & 3.49 & 0.878 & 58    \\ %4
         J1406$-$4233 & 14:06:04.79(3)   & $-$42:33:20.6(3)    & 2.4367954316(2)    & 0.681(2)    & 56846  & 2.94 & 1.56  & 58    \\ %5
         J1523$-$3235* & 15:23:37.70(3)   & $-$32:35:42.1(5)    & 1.5048345580(2)    & 0.953(9)    & 57934  & 1.25 & 1.38  & 76    \\ %6
         J1544$-$0713* & 15:44:23.527(2)  & $-$07:13:59.02(12)  & 0.484129803194(3)  & 1.13663(7)  & 57317  & 2.94 & 0.126 & 32    \\ %7
         J1604$-$3142 & 16:04:36.546(2)  & $-$31:42:06.4(2)    & 0.883890454005(11) & 0.0265(2)   & 57177  & 2.68 & 0.218 & 29    \\ %8
         J1630$-$2609* & 16:30:28.99(3)   & $-$26:09:47(3)      & 1.91241975129(15)  & 1.570(2)    & 57214  & 3.22 & 2.16  & 88    \\ %9
         J1646$-$1910* & 16:46:18.66(11)  & $-$19:10:14(8)      & 4.817735830(5)     & 10.03(14)   & 57823  & 1.03 & 4.31  & 111   \\ %10
         J1700$-$0954* & 17:00:03.228(12) & $-$09:54:43.2(8)    & 0.81731160245(4)   & 0.7520(6)   & 57199  & 3.27 & 1.15  & 39    \\ %11
         J1759$-$5505* & 17:59:34.16(2)   & $-$55:05:30.9(2)    & 0.3733921257(2)    & 0.439(2)    & 57000  & 1.03 & 0.495 & 50    \\ %12
         J1828+1221*   & 18:28:22.161(9)  & +12:21:20.9(2)      & 1.52829508058(5)   & 1.746(2)    & 57734  & 1.76 & 0.557 & 29    \\ %13
         J1910$-$0556* & 19:10:16.523(7)  & $-$05:56:29.9(4)    & 0.55760924800(3)   & 0.5961(4)   & 57325  & 2.91 & 0.811 & 42    \\ %14
         J1921$-$0510* & 19:21:29.45(2)   & $-$05:10:10.8(7)    & 0.79425387952(6)   & 0.6025(13)  & 57325  & 2.91 & 2.47  & 119   \\ %15
         J1923$-$0408* & 19:23:10.96(3)   & $-$04:08:19.4(8)    & 1.14926937103(3)   & 0.068(13)   & 57745  & 2.78 & 4.91  & 163   \\ %16
         J1928$-$0108* & 19:28:18.93(2)   & $-$01:08:55.7(4)    & 2.3657140252(4)    & 2.80(2)     & 57983  & 1.12 & 1.47  & 179   \\ %17
         J1940$-$0902* & 19:40:54.530(8)  & $-$09:02:18.0(7)    & 0.97846813844(9)   & 1.8790(14)  & 57410  & 3.25 & 1.24  & 137   \\ %18
         J1942$-$2019* & 19:42:50.34(8)   & $-$20:19:23(7)      & 0.275132723562(8)  & 0.0123(2)   & 57392  & 2.73 & 0.959 & 68    \\ %19
         J2001$-$0349* & 20:01:53.31(3)   & $-$03:49:43.5(15)   & 1.34472401604(13)  & 0.112(2)    & 57383  & 2.76 & 4.37  & 75    \\ %20
         J2136$-$5046* & 21:36:57.593(9)  & $-$50:46:56.08(10)  & 0.267322318335(10) & 0.0869(4)   & 57812  & 1.59 & 0.876 & 91    \\ \hline %21
  \end{tabular}
  \end{scriptsize}
\end{table*}

\begin{table*}
  \caption[Additional parameters for SUPERB pulsars]{Additional parameters for all pulsars, including PSR~J1421$-$4409. 
  The DM values are derived from timing or fit by \textsc{pdmp} and are quoted with 1-$\sigma$ uncertainties. 
  The \textsc{ymw16} model \citep{ymw17} could not account for the dispersion along the line of sight for PSR~J1337$-$4441 and returned the maximum distance value of 25\,kpc. 
  All RMs were derived by fitting the polarization position angle across two halves of the (256\,MHz) band. 
  Polarization fractions are given as calculated by \textsc{psrstat}, with uncertainties summed in quadrature with a 3\% error representing the systematics \citep[c.f.,][]{jk18}. 
  See Section~\ref{ssec:superb_data} for discussion of modulation indices (defined as $\sigma_{S1400}/\Bar{S}_{1400}$) and flux density distributions.  
  The radio luminosities are calculated using the \textsc{ymw16} distances with the exception of PSR~\nuller (which uses the \textsc{ne2001} value of 3.6\,kpc) and assume 15\% uncertainty on the distance. 
  The flux density and derived luminosity of PSR~\rratb are mean values including sub-threshold pulses (see Section~\ref{ssec:superb_1646}).}
  \label{tab:superb_derv}
  \centering
  \begin{threeparttable}
  \begin{scriptsize}
  \begin{tabular}{l*{10}{c}} \hline 
    Pulsar       & DM            & Distance & Distance & RM        & Linear   & Circular & Modulation & Median Flux   & Radio       & FWHM  \\
     &           & (\textsc{ymw16}) & (\textsc{ne2001}) &          & Fraction & Fraction & Index      & Density       & Luminosity  &      \\
     & (pc\,cm$^{-3}$)   & (kpc)     & (kpc)     & (rad\,m$^{-2}$) & (\%)     & (\%)     &       & ($S_{1400}$; mJy) & (mJy\,kpc$^2$) & ($W_{50}$; ms)  \\ \hline
    J0750$-$6846 & 53.8(4)       & 0.29     & 2.2      & $-$31(2)  & 41(3)    & 10(4)    & $0.4(1)$ & $1.7(2)$    & $0.14(5)$   & 188.5(9)  \\  % lum=$8^{+84}_{-8}$ NE2001
    J1115$-$0956 & 16.1(2)       & 1.1      & 0.68     & $-$9(10)  & 15(4)    & 8(4)     & $0.5(1)$   & $0.22(5)$     & $0.3(1)$   & 17(1)  \\
    J1207$-$4508 & 84.5(5)       & 2.6      & 3.1      & $-$64(5)  & 38(3)    & --       & $0.4(1)$   & $0.31(6)$     & $2.1(7)$    & 73(2)  \\
    J1244$-$1812 & 15.5(10)      & 1.1      & 0.69     & 4(5)      & 35(4)    & --       & $0.7(1)$   & $0.2(1)$     & $0.2(1)$   & 43(3)  \\
    J1337$-$4441 & 97.1(4)       & --       & 3.6      & $-$28(2)  & 27(3)    & 7(3)     & $0.45(9)$   & $1.5(1)$     & $19(6)$     & 77(1)  \\ % lum using NE2001 (>900 w/YMW16) % flux values exclude nulls
    J1406$-$4233 & 79.7(8)       & 6.8      & 2.6      & $-$68(3)  & 44(3)    & --       & $0.4(1)$   & $0.23(3)$     & $11(3)$      & 45(2)  \\
    J1421$-$4409 & 54.642(3)     & 2.1      & 1.6      & $-$21(4)  & 9(3)     & 11(3)    & $0.5(1)$ & $1.3(1)$    & $6(2)$      & 0.22(2)  \\  % RM from J1421-4409_tot_cal_eph.addTf2
    J1523$-$3235 & 72.3(9)\tnote{a} & 8.0     & 2.6     & --       & 6(4)     & 11(4)    & $0.5(2)$   & $0.20(2)$     & $13(4)$     & 93(1)  \\
    J1544$-$0713 & 30.73(12)     & 2.6      & 1.3      & 21(14)    & 12(3)    & --       & $0.31(8)$  & $0.39(5)$     & $2.7(9)$    & 6.1(4)  \\
    J1604$-$3142 & 58.3(2)       & 2.9      & 1.7      & 37(14)    & 15(3)    & 6(3)     & $0.34(7)$  & $0.37(5)$     & $3(1)$    & 7.8(9)  \\
    J1630$-$2609 & 76.6(13)      & 5.3      & 2.3      & --        & 17(4)    & 7(4)     & $0.4(1)$   & $0.42(5)$     & $12(4)$      & 52(2)  \\
    J1646$-$1910 & 55(6)         & 2.1      & 1.6      & $-$10(4)  & 38(4)    & --       & $0.6(2)$\tnote{b} & $0.08(1)$    & $0.3(1)$   & 132(5)  \\  % recalculation: $0.61\pm0.310$ & $0.268\pm0.0075$ & $1.18\pm0.36$
    J1700$-$0954 & 64.1(11)      & 0.22     & 2.3      & --        & --       & --       & $0.3(1)$   & $0.35(8)$     & $0.017(6)$  & 33.5(8)   \\  % lum=$1.1(4)$  NE2001
    J1759$-$5505 & 63.0(3)       & 3.1      & 1.7      & --        & 13(4)    & 11(4)    & $0.3(2)$   & $0.23(2)$     & $2.2(7)$    & 10.2(4)  \\
    J1828+1221   & 69.1(4)       & 3.8      & 2.9      & 83(3)     & 42(3)    & 7(3)     & $0.41(7)$  & $0.5(1)$    & $7(3)$      & 28.4(15)  \\
    J1910$-$0556 & 88.3(7)       & 3.8      & 2.7      & 75(20)    & 19(4)    & --       & $0.2(1)$   & $0.26(2)$     & $4(1)$    & 15.2(5)  \\
    J1921$-$0510 & 96.6(7)\tnote{a} & 5.6     & 3.2     & --       & 6(4)     & 8(4)     & $0.3(1)$ & $0.40(4)$     & $13(4)$      & 32.6(8)  \\
    J1923$-$0408 & 35(3)         & 1.2      & 1.5      & --        & 8(5)     & --       & $0.3(2)$   & $0.11(1)$   & $0.16(5)$   & 14.6(11)  \\
    J1928$-$0108 & 125.8(8)      & 9.7      & 4.4      & $-$15(13) & 30(4)    & --       & $0.5(1)$   & $0.24(6)$     & $23(9)$     & 65(2)  \\
    J1940$-$0902 & 42.3(5)       & 1.9      & 1.7      & --        & --       & --       & $0.2(1)$ & $0.23(2)$     & $0.8(3)$    & 18.2(9)  \\
    J1942$-$2019 & 59.9(3)       & 5.0      & 2.2      & $-$17(20) & 11(4)    & 5(3)     & $0.3(1)$ & $0.54(8)$     & $14(5)$     & 16.9(3)  \\
    J2001$-$0349 & 67(3)         & 6.0      & 2.9      & $-$79(11) & 34(4)    & 13(4)    & $0.4(2)$   & $0.28(6)$     & $10(4)$      & 83(1)  \\
    J2136$-$5046 & 23.3(3)       & 2.2      & 0.95     & 24(18)    & 11(3)    & --       & $0.6(1)$ & $0.6(2)$    & $3(1)$    & 11.0(3)  \\ \hline
  \end{tabular}
  \begin{tablenotes}
    \item[a] PSRs~J1523$-$3235 and J1921$-$0510 were also detected in the Green Bank North Celestial Cap (GBNCC) pulsar survey at 350\,MHz with DM values of 73.3(3) and 97.5(2)\,pc\,cm$^{-2}$, respectively \citep{mess+20}.  \\
    \item[b] The flux density and modulation index of PSR~J1646$-$1910 were derived from the single-pulse distribution described in Section~\ref{ssec:superb_1646}. 
  \end{tablenotes}
  \end{scriptsize}
  \end{threeparttable}
\end{table*}

\section{Individual Pulsars of interest}
\label{sec:superb_rrats}
This survey, which repeats observing pointings and is therefore more sensitive to faint scintillating sources and intermittent or nulling pulsars than previous surveys, has revealed four new pulsars with detection fraction\footnote{The detection fraction is here defined as the number of observations in which the pulsar was detected at \snr\,$>8$ divided by the total number of observations of the source. No correction has been made for the offset of the beam before the pulsar position was determined.} $\lesssim50$\,percent. 
The observation times for each source have been adjusted to minimise non-detections due to scintillation. 
Two of these sources, PSRs~\rratb and \rrata have been classified as RRATs, being only (or most significantly) detected via their single pulses. 
%This designation is tentative for PSR~\rrata, which has not yet been solved, and, as discussed below, was not accurate for PSR~\rratb (see discussion in \citealt{kml11} for comparison). 

\subsection{PSR \rratb}\label{ssec:superb_1646}
PSR~J1646$-$1910 was initially discovered in the single-pulse pipeline showing 3 pulses in 560\,seconds. It was not detected in the periodicity search, something that is not uncommon for long-period radio-emitting neutron stars with broad pulse amplitude distributions \citep{kml11}. 
Upon further observation, and the detection of more pulses, the $4.82$-s pulse period was identified, enabling folding of available data to look for any lower level periodic emission and a more in-depth assessment of the pulse amplitude distribution. 
Follow-up observations with 1800-s integrations showed the average phase-folded flux density to be sub-threshold in $> 50$\% of observations even with the deeper pointings, but bright single pulses could be seen in the majority of observations.

\begin{figure}
    \centering
    \includegraphics[trim=0 0 9.75cm 6.75cm, clip, width=\columnwidth]{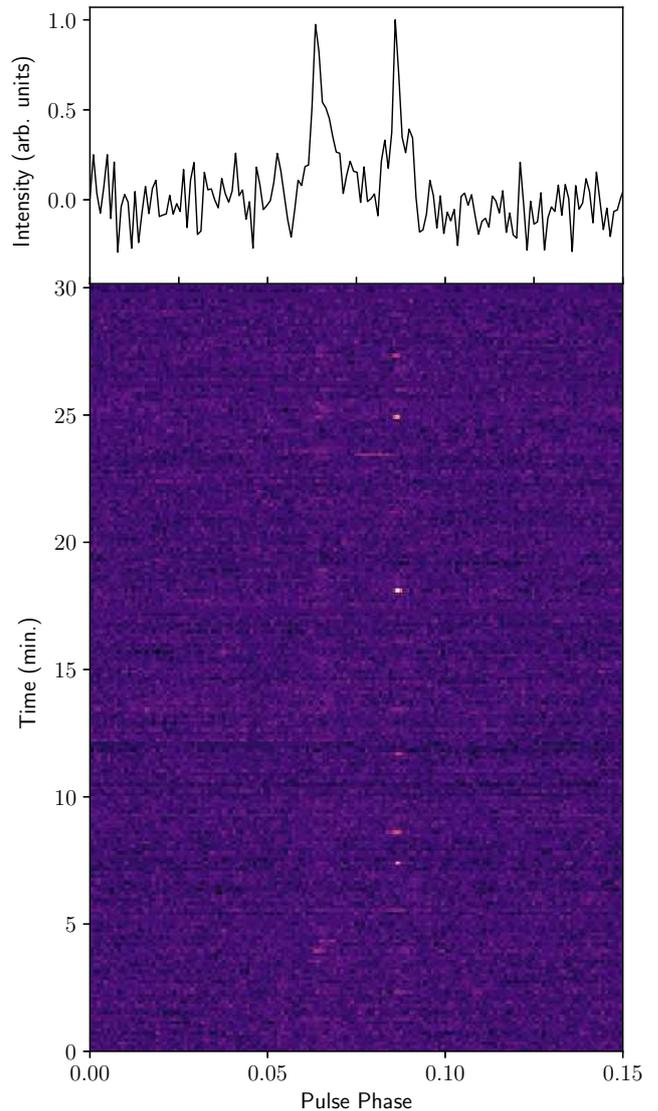}
    \caption[Single observation plot of PSR~\rratb]{Plot of PSR~\rratb showing a single observation with CASPSR, zoomed in on the peak. The bottom panel shows the intensity per 8-second sub-integration, with dark colours indicating the background noise and bright pixels showing the highly modulated pulses. We note that a given sub-integration has, on average, 1.66 periods. The top panel shows the integrated intensity of the observation.}
    \label{fig:superb_J1646_wfall}
    % python $RES/code/make_paper_plots.py -W -A J1646-1910_2018-01-21.zcgt.Fpzec -F -z 0.0,0.15
\end{figure}

\begin{figure}
    \centering
    \includegraphics[width=0.9\columnwidth]{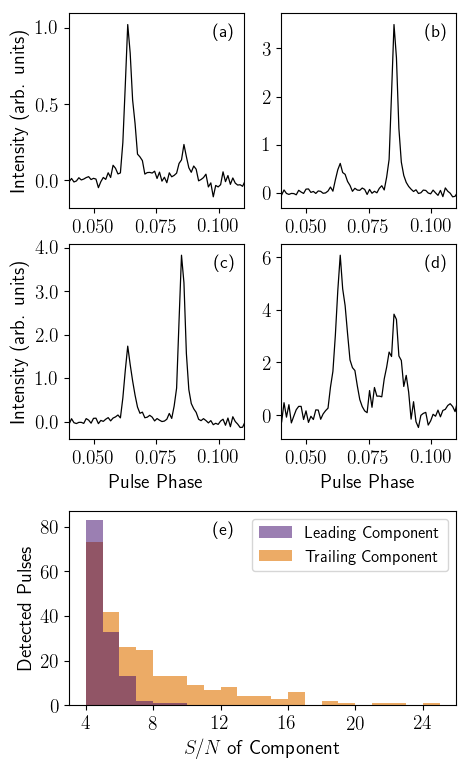}
    \caption[Profiles and component histograms for PSR~\rratb]{Integrated profiles of PSR~\rratb based on component detections in single-pulse sub-integrations. Panel (a) shows the integrated profile of 124 sub-integrations selected based on the measured \snr of the leading component, excluding those selected for the trailing component. Panel (b) shows the 230 sub-integrations selected for the trailing component, excluding those for the leading component. Panel (c) shows all 363 sub-integrations selected for either component. Panel (d) shows the opposite: 9371 sub-integrations in which neither component was above the threshold. Finally, panel (e) shows the histogram of \snr measured for each component (calculated over the phase range 0--0.15).}
    \label{fig:superb_1646_profs}
\end{figure}

Fig.~\ref{fig:superb_J1646_wfall} shows an example 30-minute observation, recorded with the CASPSR system, which has a fixed sub-integration length of 8\,s. 
This is not ideal for this pulsar, as each sub-integration contains 1.66 periods; this only minimally affects detectability but can skew emission statistics.
The \snr of the integrated profile of all BPSR observations summed together is $\approx 44$, % from pdmp, width = 132 ms
in a total of 13.0\,hours, corresponding to a phase-averaged flux density of $\approx0.08(1)$\,mJy, using the reported SEFD for the CASPSR/BPSR observing system from \citet{jvsk+18} and after correcting for the offset from the pulsar position. 
We further estimated the modulation index from the \snr values per sub-integration, calculated in the limited phase range $<0.15$, selecting only sub-integrations with $S/N > 4$ and with the peak value in the expected ranges for the two components (as described below). The modulation index, $m_{1400} = 0.6(1)$, is similar to those of the other SUPERB pulsars (listed in Table~\ref{tab:superb_derv}) but likely reflects intrinsic variations in the emission rather than effects of the interstellar medium, since we calculate this using single-pulse sub-integrations.
We note that the \snr values of single pulses shown in Fig.~\ref{fig:superb_1646_profs} imply a single-pulse flux density of up to $\approx0.22$\,Jy  for the trailing component (up to $\approx0.1$\,Jy for the leading component). % from radiometer equation: set \snr=9-24, SEFD=35 Jy, Tobs=14-23ms (no sqrt(P-W/W))

A closer inspection of the single-pulse sub-integrations from all observations of PSR~\rratb reveals the nature of the pulsar as a faint source with bright single pulses. 
We examined the single-pulse sub-integrations of the total 13.15\,hours of observations (13.0\,hours after removing RFI-contaminated sub-integrations) and identified the sub-integrations with detectable pulses according to the following criteria: 
A) $S/N > 4$ calculated with an on-pulse region in the phase range 0.059--0.07 (the leading component) or 0.08--0.093 (the trailing component) and off-pulse in the phase range 0.0--0.15, excluding the component region; 
B) width of the candidate pulse $>1$\,bin; and 
C) visual inspection to remove remaining RFI and spurious detections. 
Fig.~\ref{fig:superb_1646_profs} shows four selections of the data (out of 9734 sub-integrations, with 1\% deleted due to RFI): The top two panels show the integrated profiles for the sub-integrations selected only for the presence of each component, excluding the sub-integrations in which both components are detected. The middle-left panel shows the integrated profile for all sub-integrations with either or both components detected. 
The middle-right panel shows the remaining sub-integrations, with neither component detected above the threshold, showing both components have significant sub-threshold emission. 
In the bottom panel of Fig.~\ref{fig:superb_1646_profs}, the histograms of \snr for each component are approximately log-normal, with a larger number of bright pulses from the trailing component.  
From these statistics, assuming the emission of bright pulses is independent and random, we estimate an average of 1 pulse with $\snr>4$ every 26\,rotations ($\approx124$\,seconds).

\subsection{PSR~J1337--4441}
PSR~\nuller is a 1.3-second pulsar that is observed to null on timescales of minutes, as noted in \paperi. 
As shown in Fig.~\ref{fig:superb_polsA}, the pulsar has a double-peaked profile with significant linear and circular polarization. 
With a DM of 97.1\,pc\,cm$^{-3}$, the \textsc{ymw16} electron density model estimates a distance of $>25$\,kpc, indicating the model does not account for the observed DM along the line-of-sight (the maximum DM along the line-of-sight is given as 95.56\,pc\,cm$^{-3}$). The \textsc{ne2001} model, however, yields a distance of only 3.6\,kpc. Clearly the distance to this pulsar is highly uncertain. Nevertheless, we use this lower value when calculating the radio luminosity in Table~\ref{tab:superb_derv} but caution that it could be significantly higher.
The pulsar, when not nulled, is bright (median $S_{1400}= 1.5$\,mJy), and has a modulation index of 0.45 (after integrating to 64-second sub-integrations and excluding sub-integrations where $S_{1400} < 3\sigma_{S1400}$).

\begin{figure}
    \centering
    \includegraphics[width=0.475\textwidth]{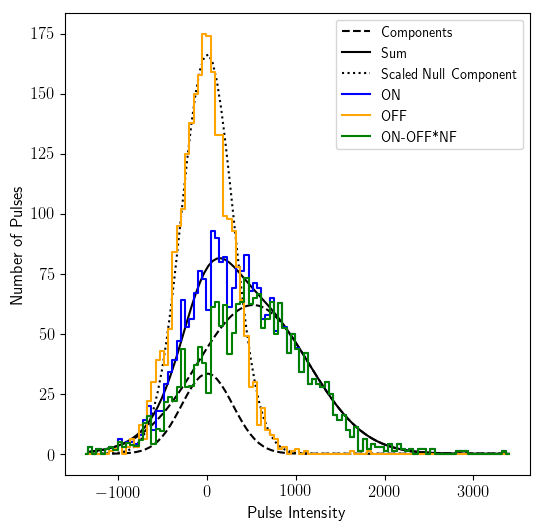}
    \caption[Intensity distribution of PSR~\nuller showing nulling]{The distribution of on- and off-pulse intensities for PSR~\nuller, fit by Gaussian functions using the \citet{ksfv18} Gaussian mixture method. The on- and off-pulse integrated intensities are plotted in blue and orange, respectively. The Gaussian components are plotted as the black dashed lines, with the solid line indicating the sum. The black dotted line shows the ``null component'' scaled by the nulling fraction. The green line indicates the difference of the on-pulse distribution and the off-pulse distribution scaled by the nulling fraction.}
    \label{fig:superb_nulldist}
\end{figure}

To determine the nulling fraction of this pulsar, we implemented the Gaussian mixture method from \citet{ksfv18}, using $\sim4.9$\,hours of BPSR observations, folded using the timing ephemeris to produce single-pulse sub-integrations, which were then integrated by a factor of 7 to improve \snr. 
The measured pulse intensities, summed over the given phase ranges, are calculated for each sub-integration for the on-pulse (phase range 0.45--0.55 as shown in Fig.~\ref{fig:superb_polsA}) and off-pulse (90$^{\circ}$ from the on-pulse region) regions. 
The Gaussian mixture method fits two Gaussian components to the on-pulse intensity distribution: one of the components models the nulls and the other the detected pulses. 
The component modelling the nulls is then scaled to the amplitude of the off-pulse intensity distribution, giving the nulling fraction as the ratio of the amplitudes. 
We show the distributions of measured pulse intensities and the Gaussian model in Fig.~\ref{fig:superb_nulldist}. 
The nulling fraction was found to be $\approx0.20$, and a Kolmogorov--Smirnov test comparing the model fit to the on-pulse intensity distribution returned a value of 0.013, corresponding to a $p$-value of 0.8, indicating no evidence to reject the null hypothesis that the data were sampled from the best-fit distribution. 
For each sub-integration, the probability of nulling was also calculated, and we use this to estimate the nulling timescale by determining numbers of successive pulses with nulling probabilities greater than 62\% of the maximum.\footnote{The maximum null probability varies depending on the \snr per sub-integration. The threshold of 62\% was selected based on fig.~1 from \citet{ksfv18}, and the results were compared with a by-eye estimation and found to be reliable.} 
We find that the null lengths are $\approx 30$ pulse periods on average (39\,seconds) and range from 7 to $>300$\,pulse periods, or a few seconds to minutes. 
The longest null confirmed by eye was approximately 8\,minutes. 
We note that the pulsar was too faint to use single-pulse sub-integrations, and the data was integrated by a factor of 7 to ensure significant signal in each sub-integration. 
Therefore, the nulling fraction is an underestimate of the true value, and the estimation of the nulling timescale is insensitive to nulls of less than 7 pulse periods. 
The lengths of the observations included in this analysis range between 270 and 1200\,s, so we are likewise insensitive to nulls of longer timescales, but the presence of emission in every available observation does not support the possibility of long nulls being common.

\subsection{Improved timing solution for PSR~J1421$-$4409}\label{ssec:superb_1421}

One of the two MSPs found in SUPERB, J1421$-$4409, is a 6.4-ms pulsar with a likely white dwarf companion in a 30.7-day orbit (\paperi). 
The pulsar has continued to be observed regularly at Parkes in a separate programme (P789, PI Spiewak), and now has approximately 3 times as many ToAs extending over a baseline of 5.1 years. 
This yields a significantly improved timing solution, in particular a significant proper motion measurement and a 4-$\sigma$ detection of the time derivative of the semi-major axis, $\dot x=1.3(3)\times10^{-13}$\,lt-s\,s$^{-1}$. 
Using the improved proper motion measurements, we re-derive the transverse velocity and account for the Shklovskii effect on the period derivative \citep{shklovskii70}, resulting in an estimated intrinsic period derivative of $6(1)\times10^{-21}$\,s\,s$^{-1}$ as compared to the observed period derivative of $1.2366(7)\times10^{-20}$\,s\,s$^{-1}$. 
We also derive an upper limit on the contribution from the orbital period derivative, $\dot P_{\rm b}$. 
Additionally, with a single observation with the UWL, we were able to significantly improve upon the previous DM value. The pulsar signal was faint across most of the 3.3\,GHz band, but we formed ToAs from 10 frequency channels from 700\,MHz to 1340\,MHz and derived the new value of 54.642(3)\,pc\,cm$^{-2}$, assuming no profile evolution across the band. 
The improved ephemeris and derived parameters are given in Table~\ref{tab:superb_j1421} (polarization parameters are provided in Table~\ref{tab:superb_derv}) and the pulse profile and position angle sweep are shown in Fig.~\ref{fig:superb_polsA}.

\begin{table}
    \centering
    \caption[Full ephemeris for PSR~J1421$-$4409]{Updated ephemeris for PSR~J1421$-$4409. 
    The DM was derived from the wideband data as described in section \ref{ssec:superb_1421}. 
    The value for the period epoch was also used for the position epoch and DM epoch. We use the `ELL1' binary model, which uses the epoch of ascending node, $T_{\rm ASC}$, and the first and second Laplace-Lagrange parameters, $\epsilon_1$ and $\epsilon_2$. The value of the derivative of the orbital period, $\dot P_{\rm b}$, is a 2-$\sigma$ upper limit.}
    \begin{scriptsize}
    \begin{tabular}{lc} \hline
        Parameter                                 & Value \\ \hline
        Right Ascension (J2000) (h:m:s)           & 14:21:20.96387(5) \\
        Declination (J2000) (d:m:s)               & $-$44:09:04.553(1) \\
        Period, $P$ (s)                           & 0.0063857288383424(1) \\
        Period derivative, $\dot P$ (s\,s$^{-1}$) & $1.2366(7)\times10^{-20}$ \\
        Period epoch (MJD)                        & 57775.4 \\
        DM (pc\,cm$^{-3}$)                        & 54.642(3) \\
        Proper motion in RA (mas yr$^{-1}$)       & $-$11.6(4) \\
        Proper motion in Dec (mas yr$^{-1}$)      & $-$7.9(8) \\
        Binary model                              & ELL1 \\
        Orbital period, $P_{\rm b}$ (d)           & 30.74645342(3) \\
        $\dot P_{\rm b}$ (s\,s$^{-1}$)            & $<2\times10^{-10}$ \\
        Semi-major axis, $x$ (lt-s)               & 12.706667(1) \\
        $\dot x$ (lt-s\,s$^{-1}$)                 & $1.3(3)\times10^{-13}$ \\
        $T_{\rm ASC}$ (MJD)                       & 57762.4168881(5) \\
        $\epsilon_1$                              & $7.6(2)\times10^{-6}$ \\
        $\epsilon_2$                              & $1.00(2)\times10^{-5}$ \\
        Units                                     & TCB \\
        Solar System Ephemeris                    & DE421 \\
        Data span (yr)                            & 5.1 \\
        Weighted RMS residual ($\upmu$s)          & 6.61 \\ 
        Number of ToAs                            & 333 \\ \hline
        \multicolumn{2}{c}{Additional Parameters} \\ \hline
        $\log(B_{\rm surf}/{\rm G})$              & 8.5 \\
%        DM-derived distance (kpc)                 & 2.1 \\
        $\mu_{\rm Tot}$ (mas yr$^{-1}$)           & 14.0(6) \\
        $V_{\rm trans}$ (km\,s$^{-1}$)            & 140(30) \\
        $\dot P_{\rm int}$ (s\,s$^{-1}$)          & $6(1)\times10^{-21}$ \\
        Eccentricity, $e$                         & $1.26(2)\times10^{-5}$ \\
        $\omega$ (deg)                            & 37.3(8) \\
        Mass function (M$_{\odot}$)               & 0.002330190(7) \\
        Min.~companion mass (M$_{\odot}$)         & 0.1757 \\
%        $W_{50}$ (at $\sim1.4$\,GHz; ms)          & 0.22(2) \\ 
        $W_{10}$ (at $\sim1.4$\,GHz; ms)          & 2.72(2) \\ \hline 
        \end{tabular}
        \end{scriptsize}
    \label{tab:superb_j1421}
\end{table}   %%%%  All checked May 22, 2020

\subsection{Unsolved objects: J0326$-$29, J0621$-$55, J1306$-$4035}
Three pulsars remain unsolved after completion of timing observations due to insufficient detections, and these are summarised in Table~\ref{tab:superb_pesky} and are described below.

\begin{figure}
    \centering
    \includegraphics[width=\columnwidth]{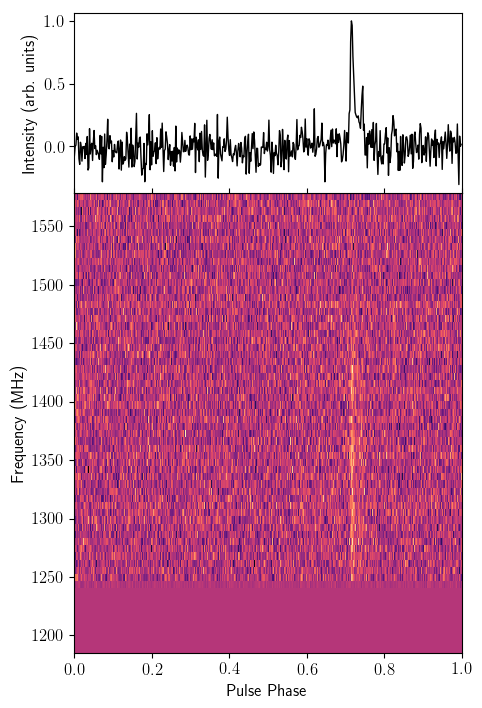}
    \caption[Detection plot of PSR~J0326$-$29]{Initial observation of PSR~J0326$-$29, de-dispersed to the optimised DM, showing the persistence of the signal across the frequency band during the 560-second observation. }
    \label{fig:superb_0326}
\end{figure}

The FFT search pipeline revealed in early 2017 a candidate with a period of 0.7679\,s and a DM of 37\,pc\,cm$^{-3}$, PSR~J0326$-$29 (see Fig.~\ref{fig:superb_0326}). 
The source was immediately re-observed three times, for a total of 32\,minutes in close succession, and was detected in each observation with \snr\,$\sim$\,10--20. 
The source was also detected in a blind search with the FFA pipeline in earlier observations.
However, in $\sim$\,15 hours of follow-up during the regular timing programme, no emission was detected, indicating the source is likely intermittent with a nulling fraction of $\gtrsim$0.9.

PSR~\rrata was discovered in a single-pulse search in early 2015 (\paperi) with a DM of 23\,pc\,cm$^{-3}$, but the period has not been determined. 
A total of 6 pulses (with \snr\,$\sim$\,10--30) have been detected at the position of PSR~\rrata with similar widths and DMs, out of a total of $\approx11.5$\,hours, with no pulses detected with \snr\,$<$\,10 (above the threshold of 6.5). 
The minimum separation of the detected pulses is $\sim150$\,seconds.

\begin{table}
    \caption[Basic parameters of unsolved SUPERB sources]{Properties of currently-unsolved sources from original discoveries. The period and DM values and uncertainties are the best-fit parameters from \textsc{pdmp}. }
    \label{tab:superb_pesky}
    \centering
    \begin{scriptsize}
    \begin{tabular}{lcccc} \hline 
        Pulsar     & RA & Declination    & Period     & DM  \\
                   & (J2000; h:m:s)  & (J2000; d:m:s) & (ms) & (pc\,cm$^{-3}$) \\ \hline
        J0326$-$29 & 03:26(2)        & $-$28:50(7)    & 767.903(6) & 32(4) \\
        \rrata     & 06:22(2)        & $-$55:57(7)    & --         & $\sim$23  \\ 
        \troll     & 13:06:56.0(5)   & $-$40:35:23(7) & 2.20453(2) & 34.95(6) \\\hline 
    \end{tabular}
    \end{scriptsize}
\end{table}

In addition, PSR~\troll, a redback originally published as J1306$-$40 in \paperi, has not been solved. 
The pulsar has a rotational period (from the discovery observations) of 2.20453\,ms and a DM of $34.95$\,pc\,cm$^{-3}$, and an orbital period of 26.3\,hr (\paperi), which was confirmed via its optical and X-ray counterparts \citealt{linares17}). 
Recently, \citet{sscs19} analysed optical data of the source and determined additional orbital parameters, including the minimum mass of the neutron star, $1.75\,{\rm M}_\odot$. 
The 4 detections reported in \paperi correspond to orbital phases 0.6--0.8 based on the orbital parameters from \citet{sscs19}. 
In this convention, the pulsar is on the far side at orbital phase 0.25.
The source has been observed numerous times as part of the P892 and P789 programmes, but preliminary searches of the data have been largely unfruitful. 
An observing programme targeting optimal orbital phases when the pulsar is on the near side of the companion would help increase the detection fraction.

%%%%%%%%%%%%%%%%%%%%%%%%%%%%%%%%%%%%%%%%%%%%%%%%%%%%%

\section{Discussion and Conclusions}
\label{sec:superb_disc}
The SUPERB programme has discovered 27 pulsars to date. 
%Our pulsars are the best and most beautiful. 
The repetition of observations at the same position enables us to detect a significant number of RRATs, nulling pulsars, and intermittent pulsars, and faint pulsars with large modulation indices, like PSR~\rratb. 
Continued searches of the data could potentially reveal several more such sources, in addition to ordinary pulsars in the remaining data.

\subsection{Polarization}
Figures~\ref{fig:superb_polsA}--\ref{fig:superb_polsE} show the calibrated profiles and polarization position angle measurements for 23 pulsars. 
The polarization fractions (listed in Table~\ref{tab:superb_derv}) are consistent with those found for similar pulsars. 
In particular, \citet{jk18} have examined the correlation between the spin-down luminosities, $\dot E$, and fraction of linear polarization for $>500$\,pulsars, including young pulsars. 
They find, in agreement with \citet{wj08}, that the linear polarization fractions of low-$\dot E$ pulsars generally do not exceed 55\%, with few exceptions. 
The SUPERB pulsars, with $\dot E$ values ranging from $6.1\times10^{29}$ to $4\times10^{32}$\,erg\,s$^{-1}$, further confirm this apparent upper limit on the linear polarization distribution, with the highest measured value of 44\%.

Additionally, 16 pulsars have Rotation Measures (RMs) determined using \textsc{rmfit} to fit the polarization position angle across the 256\,MHz band of the DFB4 data (the fit for PSR~J1421$-$4409 used the 320\,MHz CASPSR data). The values were checked using the \textsc{rmfit} brute force method to find the peak linear polarization fraction as a function of RM. 
The measured RMs are not unusually high, with the largest RM/DM for PSR~J1828+1221 corresponding to a line-of-sight average magnetic field strength of 1.47\,$\upmu$G.

\subsection{Detections in Other Surveys}
Two of the pulsars discovered in SUPERB, PSRs~J1523$-$3235 and J1921$-$0510, were detected in the GBNCC pulsar survey (\citealt{mess+20}; listed as J1524$-$33 and J1921$-$05B therein), as noted in Table~\ref{tab:superb_derv}. 
The GBNCC survey \citep{slr+14} uses a central frequency of 350\,MHz, thereby allowing a more precise measurement than ours of the DMs. 
For PSRs~J1523$-$3235 and J1921$-$0510, the DMs resulting from the GBNCC survey are 73.3(3) and 97.5(2)\,pc\,cm$^{-2}$, respectively, consistent at the $\sim$1-$\sigma$ level with our measurements at 1400\,MHz. 
Also, by using the radiometer equation to estimate the flux density from the measured \snr of the detections, \citet{mess+20} derive flux density values ($S_{350}$) of 2.0(5) and 1.8(4)\,mJy, respectively. 
Combined with our $S_{1400}$ measurements and assuming standard power-law spectra, the estimated spectral indices of PSRs~J1523$-$3235 and J1921$-$0510 read $-1.7\pm0.9$ and $-1.1\pm1.3$, both of which are consistent with the average power-law spectral index of the pulsar population ($\approx-1.6$; \citealt{jvsk+18}).

We also retrieved archival data from the HTRU hilat survey near the positions of our discoveries and folded the closest observations (within $\sim12$\,arcmin) using our finalised ephemerides. 
Of the 69 observations within 12\,arcmin of the pulsar positions\footnote{We excluded the unsolved sources from our search for HTRU hilat data, and PSR~J1828+1221 was outside the region observed by HTRU, leaving 21 pulsars with hilat observations within 12\,arcmin.}, 47 were available to us for folding\footnote{The remaining 22 hilat observations were not accessible at the time of this analysis.}. 
With a \snr threshold of 8, we detected 9 SUPERB pulsars in the available hilat data: PSRs~J1115$-$0956, J1337$-$4441 (2 detections), J1421$-$4409, J1544$-$0713 (2 detections), J1646$-$1910, J1700$-$0954, J1759$-$5505, J1921$-$0510, and J1942$-$2019. 
We note that the detections of PSRs~J1700$-$0954 and J1759$-$5505 had $S/N\sim7.5$, but the pulses appeared consistent with the width and shape of the signals from SUPERB data, and we then consider those genuine detections. 
We further note that a third observation of J1544$-$0713 appeared to have a signal with the correct characteristics but had $S/N=6.6$.

In order to address the reasons for the non-detection of these pulsars in the HTRU hilat survey, we note that the latter survey has not been searched with an FFA so far \citep{mbc+19}, and that 6 out of the 9 pulsars above were indeed initially detected in SUPERB by using an FFA search. 
There are two likely causes to the superior sensitivity of the FFA in the SUPERB survey processing: firstly that the method is fully phase-coherent, as opposed to the standard FFT method that employs incoherent harmonic summing; secondly, that the low-frequency noise mitigation algorithms routinely employed in FFT search codes (so-called ``spectral whitening'') have been shown to significantly reduce the sensitivity to longer period ($P \gtrsim 1$\,s) pulsars \citep[e.g.,][]{lbh+15,cbckz17,vhkr17,pkr+18}. 
An in-depth theoretical comparison between the sensitivity of the FFA and the standard FFT method will be published soon \citep[][submitted]{mbskl20} along with an improved version of the FFA software that was used in the first SUPERB FFA search. Thus, a new search of the SUPERB data with this new software should enhance even more the sensitivity to long period and short duty cycle pulsars.

%%%%%%%%%%%%%%%%%%%%%%%%%%%%%%%%%%%%%%%%%%
\section*{Data Availability}
Data obtained in Parkes observing programmes P858, P892, and P789 are archived for long-term storage on the CASS/ANDS data server\footnote{\url{https://data.csiro.au/dap/public/atnf/pulsarSearch.zul}}. 
These data are publicly available 18 months from the day they are recorded.
The data are recorded in \textsc{sigproc} filterbank format, a de facto standard for pulsar search data, and are converted to
\textsc{psrfits} format for upload to the data server.
Some reduced data products (pulsar ephemerides, ToAs, and polarization profiles) can be accessed directly from the Zenodo repository at \url{https://doi.org/10.5281/zenodo.3900980}.

%%%%%%%%%%%%%%%%%%%%%%%%%%%%%%%%%%%%%%%%%%

\section*{Acknowledgements}
The authors thank the anonymous referee and scientific editor for their detailed comments that significantly improved this manuscript. 
The authors acknowledge colleagues who contributed to the completion of the SUPERB survey and indirectly to this project, including Ramesh Bhat, Manisha Caleb, Wael Farah, and Andrew Jameson. 
This research was funded partially by the Australian Government through the Australian Research Council, grants CE170100004 (OzGrav) and FL150100148. 
FJ and VM acknowledge funding from the European Research Council (ERC) under the European Union's Horizon 2020 research and innovation programme (grant agreement No.~694745). 
The Parkes radio telescope is part of the Australia Telescope National Facility which is funded by the Australian Government for operation as a National Facility managed by CSIRO. 
This work used the gSTAR national facility which is funded by Swinburne and the Australian Government's Education Investment Fund. 
This work also used the OzSTAR national facility at Swinburne University of Technology. 
OzSTAR is funded by Swinburne University of Technology and the National Collaborative Research Infrastructure Strategy (NCRIS). 
This research made use of Astropy, a community-developed core Python package for Astronomy \citep{astropy13,astropy18}, the Scipy package \citep{scipy}, and the Matplotlib package \citep{hunter07}.

%%%%%%%%%%%%%%%%%%%%%%%%%%%%%%%%%%%%%%%%%%%%%%%%%%

%%%%%%%%%%%%%%%%%%%% REFERENCES %%%%%%%%%%%%%%%%%%

\bibliographystyle{mnras}
\bibliography{superb_psr}

%%%%%%%%%%%%%%%%%%%%%%%%%%%%%%%%%%%%%%%%%%%%%%%%%%

\appendix % for the profile plots
\section{Integrated Pulse Profiles}
We show the integrated profiles for each of the 23 solved pulsars with polarization properties including the polarization position angle (P.A.). 
All profiles except that of PSR~J1421$-$4409 are from incoherently de-dispersed DFB4 data, but the DM smearing in any channel is negligible (PSR~J1421$-$4409 was observed with coherent de-dispersion with CASPSR).

\begin{figure*}
    \centering
    \includegraphics[width=0.95\textwidth]{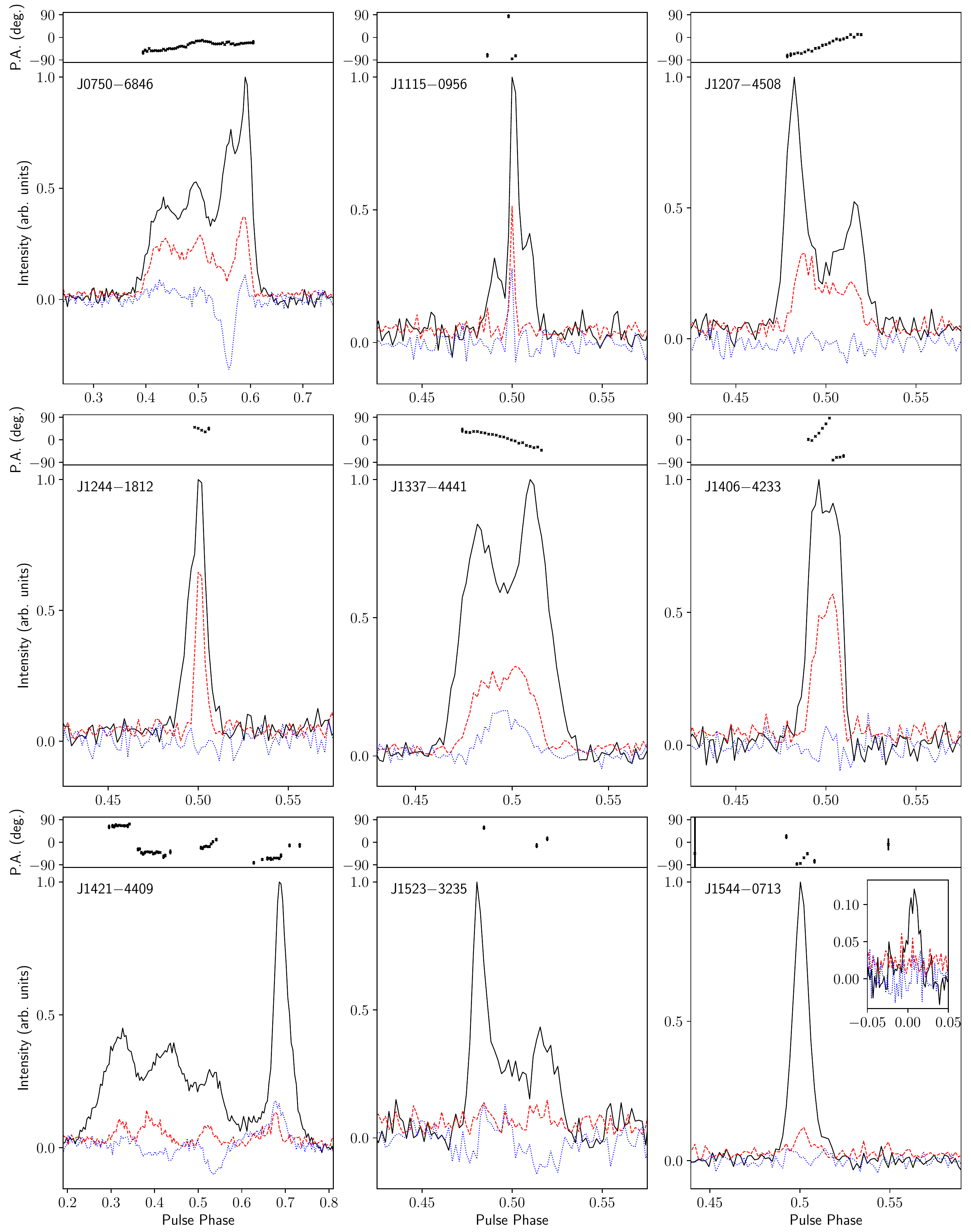}
    \caption[Polarization-calibrated profiles of a selection of SUPERB pulsars]{Polarization-calibrated profiles of PSRs~J0750$-$6846, J1115$-$0956, J1207$-$4508, J1244$-$1812, \nuller, J1406$-$4233, J1421$-$4409, J1523$-$3235, and J1544$-$0713. The bottom panel of each subfigure shows the pulse profiles, with total intensity in black, linear polarization in red, and circular polarization in blue. The top panels show the polarization angle as a function of pulse phase. DM smearing due to incoherent de-dispersion (for all profiles except PSR~J1421$-$4409 which is coherently de-dispersed) is negligible. The interpulse for PSR~J1544$-$0713, which appears half a rotation from the main pulse, is shown in the inset plot of the last panel. The number of phase bins is 512 for all pulsars shown except PSRs~J0750$-$6846 and J1421$-$4409, which use 256 bins (also note larger phase ranges shown for these pulsars). }
    \label{fig:superb_polsA}
\end{figure*}

\begin{figure*}
    \centering
    \includegraphics[width=0.95\textwidth]{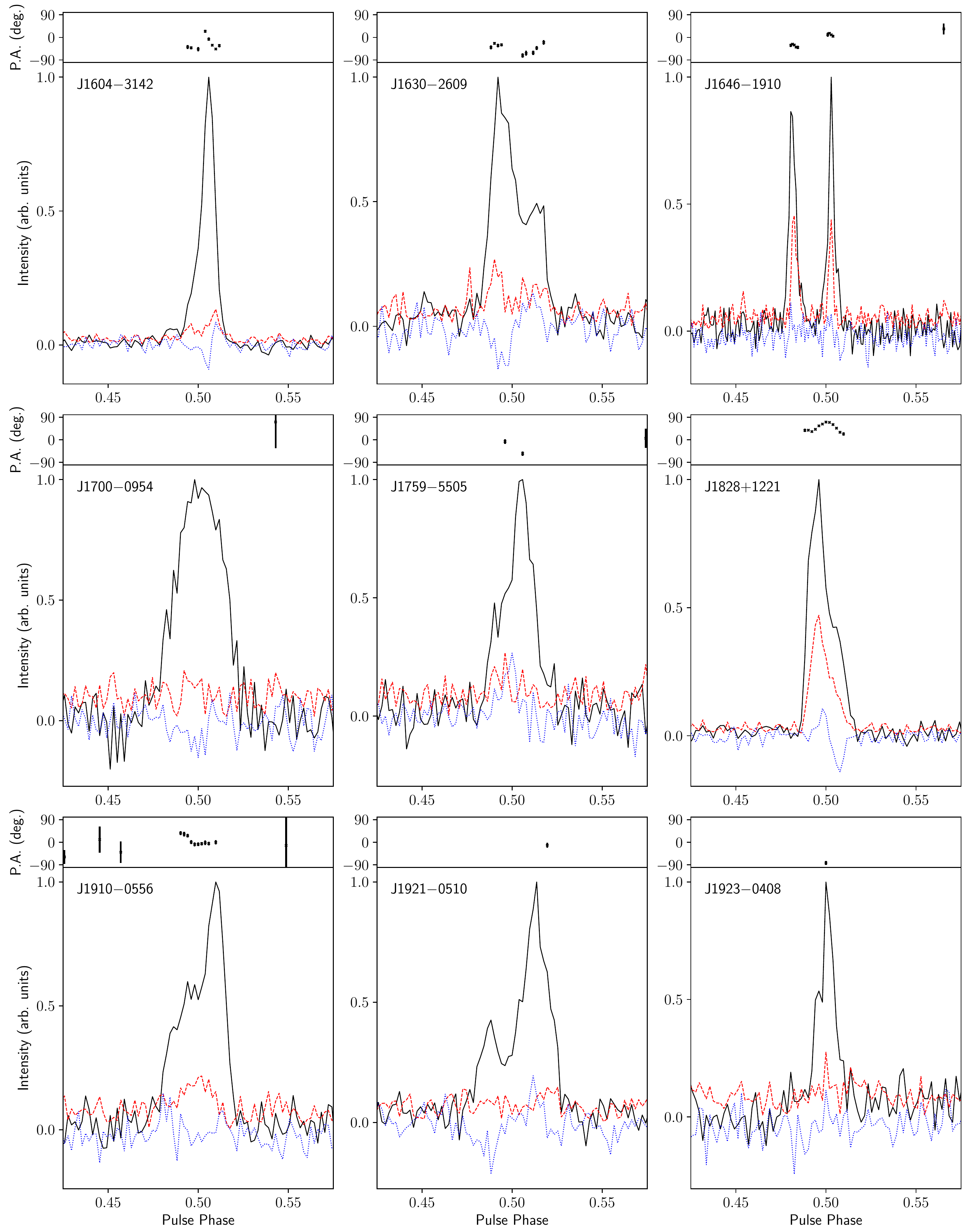}
    \caption[Polarization-calibrated profiles of additional SUPERB pulsars]{As Fig.~\ref{fig:superb_polsA}, for PSRs~J1604$-$3142, J1630$-$2609, J1646$-$1910, J1700$-$0954, J1759$-$5505, J1828+1221, J1910$-$0556, J1921$-$0510, and J1923$-$0408. The number of phase bins is 512 for all pulsars shown except PSR~J1646$-$1910, which uses 1024 bins.}
    \label{fig:superb_polsD} % careful with labels
\end{figure*}

\begin{figure*}
    \centering
    \includegraphics[width=0.95\textwidth]{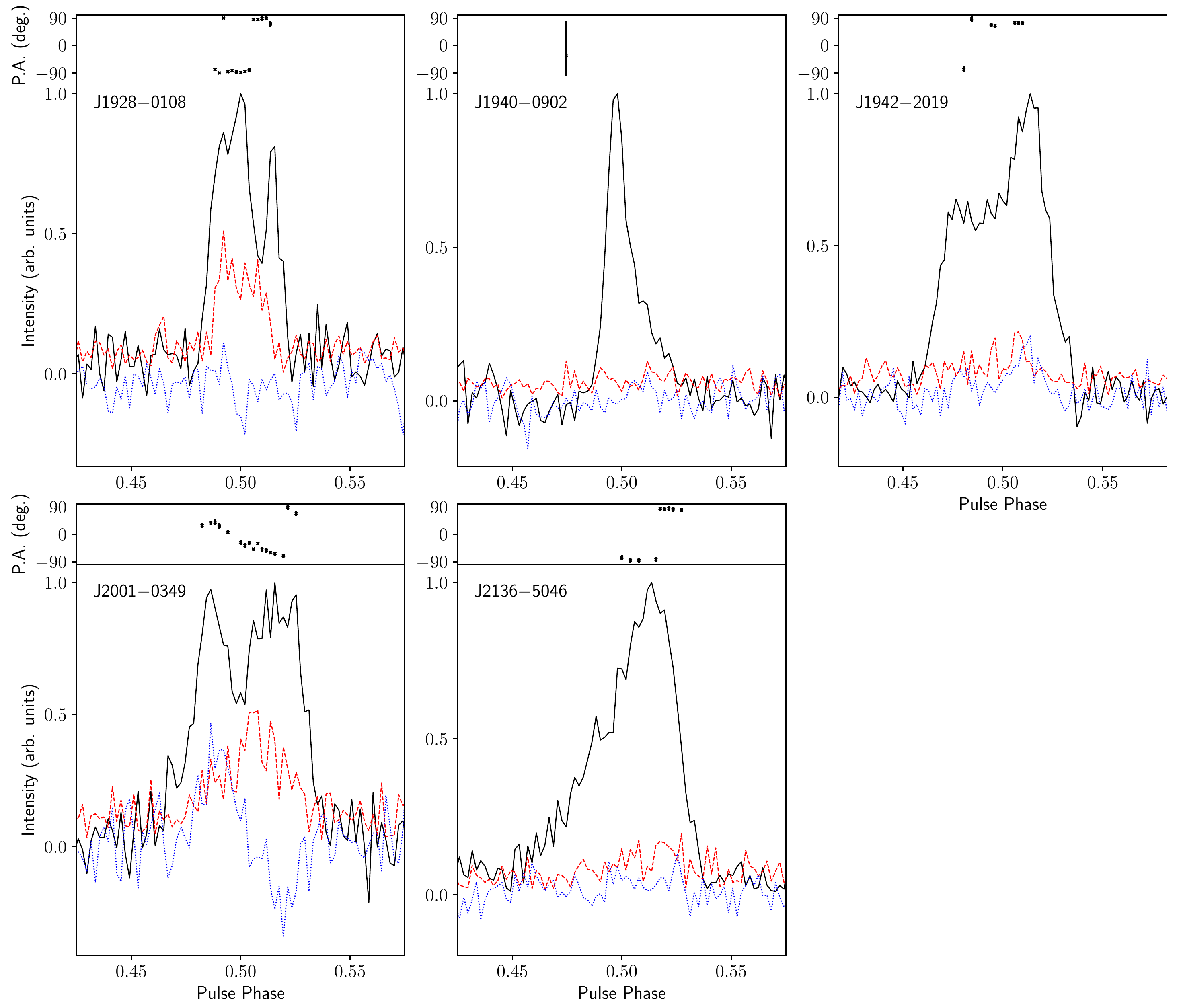}
    \caption[Polarization-calibrated profiles of additional SUPERB pulsars]{As Fig.~\ref{fig:superb_polsA}, for PSRs~J1928$-$0108, J1940$-$0902, J1942$-$2019, J2001$-$0349, and J2136$-$5046. The number of phase bins is 512 for all pulsars shown.}
    \label{fig:superb_polsE} % careful with labels 
\end{figure*}

% Don't change these lines
\bsp	% typesetting comment
\label{lastpage}
\end{document}